\documentclass[pra,showpacs,twocolumn,superscriptaddress,longbibliography]{revtex4-1}

\usepackage{amsmath}
\usepackage{amssymb}
\usepackage{graphicx}
\usepackage{dcolumn}
\usepackage{bm}
\usepackage{amsfonts}
\usepackage{subfigure}
\usepackage{float}
\usepackage{color}
\usepackage{xcolor}
\usepackage{braket}
\usepackage{hyperref}
\usepackage{comment}

\makeatletter
\def\maketitle{
\@author@finish
\title@column\titleblock@produce
\suppressfloats[t]}
\makeatother

\hypersetup{pdfborder = {0 0 0}}

\begin{document}
\title{Observation of self-patterned defect formation in atomic superfluids \\ -- from ring dark solitons to vortex dipole necklaces}

\author{Hikaru Tamura}
\email{tamurah@ims.ac.jp}
\affiliation{Department of Physics and Astronomy, Purdue University, West Lafayette, IN 47907, USA}
\author{Cheng-An Chen}
\thanks{Current address: Atom Computing, Boulder, CO 80301, USA}
\affiliation{Department of Physics and Astronomy, Purdue University, West Lafayette, IN 47907, USA}
\author{Chen-Lung Hung}
\email{clhung@purdue.edu}
\affiliation{Department of Physics and Astronomy, Purdue University, West Lafayette, IN 47907, USA}
\affiliation{Purdue Quantum Science and Engineering Institute, Purdue University, West Lafayette, IN 47907, USA}
\date{\today }

\begin{abstract} 
Unveiling nonequilibrium dynamics of solitonic and topological defect structures in a multidimensional nonlinear medium is a current frontier across diverse fields. One of the quintessential objects is a ring dark soliton (RDS), whose dynamics are expected to display remarkable interplay between symmetry and self-patterned topological defect formation from a transverse (snake) instability, but it has thus far evaded full experimental observations. Here, we report an experimental realization of RDS generation in a two-dimensional atomic superfluid trapped in a circular box. By quenching the confining box potential, we observe an RDS emitted from the edge and its peculiar signature in the radial motion. 
As an RDS evolves, we observe transverse modulations at discrete azimuthal angles, which clearly result in a patterned formation of a circular vortex dipole array. Through collisions of the vortex dipoles with the box trap, we observe vortex unbinding, vortex pinning to the edge, and emission of rarefaction pulses. Our box-quench protocol opens a new way to study multidimensional dark solitons, structured formation of topological defects, and potentially the dynamics of ordered quantum vortex matter.
\end{abstract}
\maketitle
Vortices and dark solitons are fundamental defect structures that appear in nonlinear physics at all scales, from superfluids and nonlinear optics to cosmic fluids. They play critical roles in understanding the dynamics and microscopic characteristics of the hosting medium. A quantized vortex emerges as a result of a topologically protected singularity with a $2\pi$ phase winding. 
In quantum gases, beginning with seminal experiments with dynamical optical imprinting techniques \cite{matthews1999vortices}, vortices are also produced by injecting angular momentum through stirring \cite{madison2000vortex,abo2001observation,kwon2015periodic,seo2017observation,gauthier2019giant,johnstone2019evolution}.
Several other techniques have been discovered \cite{bland2023vortices}. While most experiments have excited disordered vortices with equal or both circulations or a vortex lattice of the same charges \cite{madison2000vortex,abo2001observation}, few-vortex structures with engineered flow patterns were realized only recently \cite{kwon2021sound,hernandez2023universality}.
A dark soliton, on the other hand, features a phase jump across a nontopological defect in the wave function, and is discovered primarily through phase \cite{burger1999dark, denschlag2000generating, ku2016cascade}, density \cite{dutton2001observation,Ginsberg2005observation}, or state \cite{anderson2001watching, becker2008oscillations} engineering techniques or by matter-wave interference \cite{weller2008experimental, shomroni2009evidence,chang2008formation}.
By driving a quantum gas through a continuous phase transition, both vortices and solitonic defects are found to form spontaneously via the Kibble-Zurek mechanism \cite{weiler2008spontaneous,freilich2010real,lamporesi2013spontaneous,chomaz2015emergence,ko2019kibble}, indicating their complementary roles in a universal defect formation process.

Remarkably, in two or three dimensions, dark solitons are fundamentally connected to highly ordered vortex states of complex phase patterns through an intrinsic instability \cite{kevrekidis2004pattern}, where a self-amplifying transverse modulation can fragment a stripe (or plane) of phase defect into an ordered array of vortex and antivortex (line or ring) pairs. This fascinating process, called transverse instability (TI), has been under heavy investigation in diverse fields for decades \cite{kivshar2000self}, including quantum gas experiments \cite{anderson2001watching,dutton2001observation,shomroni2009evidence,becker2013inelastic,donadello2014observation,ku2016cascade}. 
In previous experimental studies, however, vortices were often observed as disordered decay products of dark solitons. Self-patterned, ordered vortex dipole arrays have never been clearly visualized.

Controlling soliton generations and its instability could open a doorway towards forming complex vortex structures that are arduous to reach artificially. In a two-dimensional (2D) quantum fluid, an interesting example emerges from a ring dark soliton (RDS) \cite{kivshar1994ring} that manifests as a circular dark stripe formed under rotational symmetry. 
An RDS does not disperse due to the balance between self-defocusing and wave dispersion \cite{kivshar1998dark,frantzeskakis2010dark}, similarly to straight counterparts, and it naturally exhibits radial oscillations while varying its profile.
Breaking the rotational symmetry of an RDS feeds TI \cite{kivshar2000self,toikka2013snake}. This results in elusive formation of a vortex dipole ``necklace'', which consists of a circular array of vortex-antivortex pairs \cite{theocharis2003ring}. Remarkably, such ordered vortices with alternating charges may exhibit a variety of many-body dynamics, including persistent revivals of structures \cite{pietil2006stability,shomroni2009evidence,toikka2013snake} and clusterization \cite{crasovan2002globally,mott2005stationary,pietil2006stability,cawte2021neutral}, which do not occur in disordered vortex matter \cite{gauthier2019giant, johnstone2019evolution}. Moreover, structured vortex matter can melt under significant perturbation and may eventually lead to chaos or turbulence \cite{novikov1975dynamics,toikka2013creation,cawte2021neutral}.

While RDS-like dark waves were previously engineered via phase imprinting in nonlinear optics \cite{baluschev1995generation,dreischuh2002ring} or have emerged from shock-wave emissions in atomic or polaritonic condensates \cite{hoefer2006dispersive,dominici2015real} and optics \cite{wan2007dispersive}, one central question concerning this study is whether self-patterned solitonic and topological defect formation via an RDS can be controlled and clearly observed.
Here, we show that a box-confined superfluid serves as a perfect arena \cite{navon2021quantum}. RDS formation can be realized in a box trap with a sharp wall, whose width is comparable to or smaller than the superfluid healing length. The edge profile of a superfluid can be viewed as a density defect [Fig.~1(a), top panel]. A quench-down of the potential height or an interaction quench-up would effectively cause shrinkage of the defect because the edge of the superfluid expands outwards (bottom panel). This dynamics forces the edge to emit dark solitons to conserve atom number, an effect that has recently been discussed in a case of an interaction quench-up with a perfect wall~\cite{halperin2020quench}. The mechanism is similar to an interaction quench that splits a full dark soliton as described in~Refs. \cite{gamayun2015fate, franchini2015universal}. An alternative interpretation of this edge effect is self-interference~\cite{ruostekoski2001interference}, where an expanding superfluid bounces off the wall, and the interference between the bulk and the reflected flow induces phase slips, thus forming dark solitons. This effect should occur in quenched nonlinear systems with sharp boundaries---a $(D-1)$-dimensional shell wave could form from a $D$-dimensional system, which is difficult to achieve with existing engineering techniques \cite{burger1999dark, denschlag2000generating, ku2016cascade,dutton2001observation,Ginsberg2005observation,anderson2001watching, becker2008oscillations,weller2008experimental, shomroni2009evidence,chang2008formation}.

\begin{figure}[t]
\centering
\includegraphics[width=0.5\textwidth]{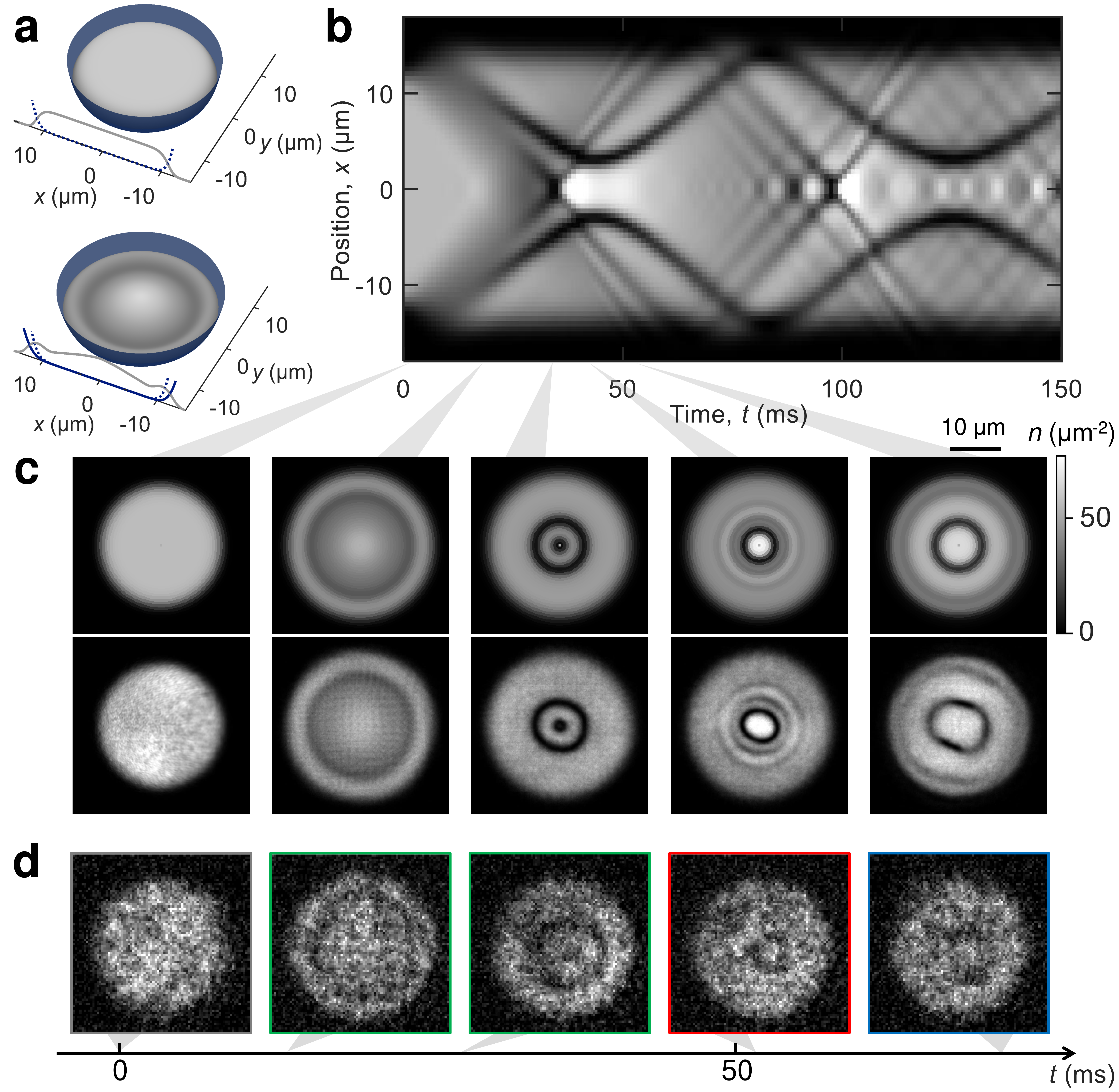}
\caption{Spontaneous formation of ring dark solitons. (a) Superfluid confined in a 2D circular box with a Gaussian wall (top) subject to a potential quench-down at $t=0$, which emits RDSs from the edge (bottom). (b) Time evolution of the density line cut across the box center and (c) 2D density images (top row), evaluated using a Gross-Pitaevskii equation (GPE). Images on the bottom row are obtained at the same indicated time, but with initial density fluctuations simulated in the GPE calculation (Appendix~\ref{App:GPE}). Single-shot in situ images in panel (d) demonstrate the formation of RDSs (green boxes), the onset of TI (red box), and the formation of vortex dipoles (blue box), respectively. Image resolution is approximately $0.8~\mu$m.
}
\label{fig:fig1}
\end{figure}

\begin{figure}[t]
\centering
\includegraphics[width=0.5\textwidth]{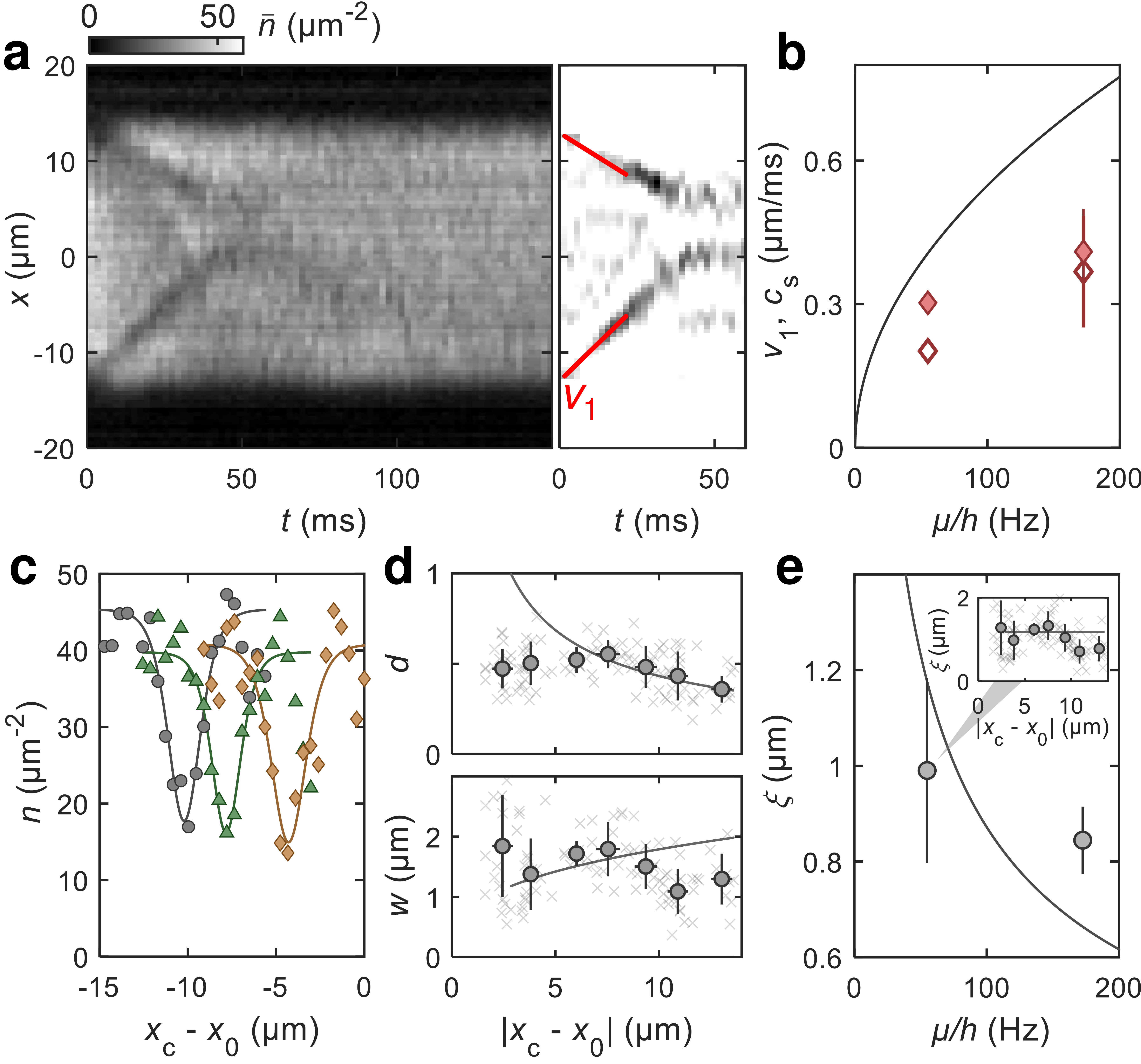}
\caption{
Characterization of ring dark solitons. (a) Left panel: Time evolution of mean density line cuts $\bar{ n}(x,\, y_0)$. Right: Center of the dark waves and linear fits (color lines) highlighted in a filtered image. (b) Propagation speeds of the darker waves determined at $x<0$ (filled symbols) and $x>0$ (open symbols), respectively. Results obtained at a higher chemical potential are plotted for comparison. The solid curve is the calculated sound speed $v_\mathrm{s}$. (c) Single-shot density line cuts at $t=17.5$ (circles), $23.5$ (triangles), and $39.5$ (diamonds)~ms, respectively. Solid lines are fits. (d) Fitted depths $d$ and widths $w$ versus radial position from single shots (crosses) and their means (circles). Solid lines are the case of a nonperturbed RDS, expected from Eq.~(\ref{eq:depth}). 
(e) Healing length $\xi=w\sqrt{d}$ determined from single-shot fit results (insets). Solid lines are expectations $\xi=\hbar/\sqrt{m\mu}$. Error bars are standard deviations.}
\label{fig:fig2}
\end{figure}

In this article, we report the first observation of self-patterned defect formation in a box-confined 2D superfluid. We demonstrate spontaneous RDS formation and unveil its radial dynamics with a symmetry-breaking TI at discrete azimuthal angles. We visualize structured fragmentation of an RDS into a necklace of vortex dipoles. The observed vortex dipole structures include not only weakly bound vortex-antivortex pairs but also coalesced vortex cores and rarefaction pulses. They are subject to collisions, interactions with the boundary, and annihilation, potentially showing rich nonequilibrium dynamics of quantized 2D vortex matter.

Our experimental scheme is illustrated in Fig.~\ref{fig:fig1}(a). A 2D circular box is enclosed by a ring-shaped repulsive wall that has an approximate Gaussian radial profile ($1/e^2$ width $\sim 5$ $\mu$m). The box confines a homogeneous 2D superfluid, with negligible thermal components, formed by cesium atoms with an initial bulk density $n \approx 50\,\rm{\mu m^{-2}}$ and prepared at a fixed coupling constant $g\approx 0.017$, which leads to a long healing length $\xi = 1/\sqrt{n g} \approx 1.2~\mu$m convenient for in situ defect measurements (Appendix~\ref{App:prep}). The chemical potential is $\mu \approx \hbar^2 /(m \xi^2) \approx k_{B} \times 3~$nK, where $\hbar=h/2\pi$ is the reduced Planck constant, $m$ is the atomic mass, and $k_{B}$ is the Boltzmann constant. At time $t=0$, the height of the wall potential is quenched from $k_{B} \times 35\,\rm{nK}$ to $9\,\rm{nK}$. Because of the much-reduced repulsion from the Gaussian wall, the superfluid would expand outwards, forcing the boundary to emit a ring-shaped dark wave.

One signature of an RDS is its radial collapse dynamics, described by a wave function that is essentially identical to a 1D dark soliton in the radial coordinate \cite{kivshar1994ring}, $\psi(r,t) = \sqrt{n}\left( i\sqrt{1-d} + \sqrt{d} \tanh \frac{r-r_\mathrm{c}}{w} \right)e^{-i\mu t/\hbar}$, where $r_\mathrm{c}(t)$ is a time-dependent radius. The depth $d$ controls the radial velocity $\dot{r}_\mathrm{c} = \pm v_\mathrm{s}\sqrt{1-d}$ and the characteristic width $w=\xi/\sqrt{d}$, where $v_\mathrm{s}=\hbar/m\xi$ is the sound speed. Unlike linear dark solitons, the depth of an RDS does not remain constant but acquires an adiabatic radial dependence to conserve its energy (Appendix~\ref{App:radial}),
\begin{equation}
    d \approx d(t_{i}) \left[ \frac{r_\mathrm{c}(t_{i})}{r_\mathrm{c}} \right]^{2/3} \, ,  \label{eq:depth}
\end{equation}
where $d(t_{i})$ and $r_\mathrm{c}(t_{i})$ are the initial conditions. Both radial speed and width also pick up their radial dependences accordingly. For a shrinking RDS, the maximum depth $(d=1)$ can be reached at a minimum radius $r_{\mathrm{min}} = r_\mathrm{c}(t_{i}) d(t_{i})^{3/2} \gtrsim \xi$. At this point, the radial motion would come to a complete stop, followed by expansion \cite{kivshar1994ring}. For a shallower or smaller RDS with $r_\mathrm{c}(t_{i}) d(t_{i})^{3/2} \lesssim \xi$, it could collapse into a single defect.

Numerical evaluation of a GPE (Appendix~\ref{App:GPE}) supports RDS emission from this quench protocol. As shown in Figs.~\ref{fig:fig1}(b) and~\ref{fig:fig1}(c), initially two distinct RDSs can be seen to emerge from the edge of the wave function. A slower-moving, darker ring shrinks until it reaches the maximum depth and a minimum radius. The ring then rebounds and expands radially. Another shallower, faster-moving ring appears to collapse at the center but would emerge again as an expanding ring. Both RDSs are later reflected off the box wall as discussed in the case of 1D solitons \cite{sciacca2017matter}, exhibiting bouncing dynamics periodically. These box-trapped RDSs cross each other multiple times with preserved shapes, and appear to be long-lived if rotational symmetry is not explicitly broken. However, they radiate additional shallow RDSs after collapsing at the box center, when the radial motion becomes nonadiabatic.

We experimentally confirm RDS emission from in situ images of box-quenched superfluids. Figure~\ref{fig:fig1}(d) shows qualitative correspondences between single-shot experiment density profiles and GPE results. A prominent dark ring is clearly visible within $t \lesssim 50~$ms until a minimum radius is reached. 
The line cut density [Fig.~\ref{fig:fig2}(a)] averaged over different experimental shots clearly shows the dark ring's radial bouncing dynamics. A less visible, shallower dark wave is found to cross near the box center at $t\approx 30~$ms, similar to the GPE result [Fig.~\ref{fig:fig1}(b)]. 

Initial radial velocities of the darker ($v_1$) rings are shown in Fig.~\ref{fig:fig2}(b). Wave speeds from samples with a larger chemical potential, but with the same quench protocol, are plotted for comparison. All measured velocities are significantly lower than the sound speed. We note that there is an anisotropy in the observed wave velocities across the box center. 
This originated from an azimuthal variation in the wall width, which is due to aberration in our optical potential, and this gives a slight anisotropy in the soliton depth and radial velocity as well. As a result, the dark ring center appears to be drifting slightly in the box, with $(x_{0},\,y_{0})\approx (2.6,\,-2.2)\,{\rm{\mu m}}$ at $t \approx 50~$ms. For all quoted positions in the following analyses, the shift has been corrected.

The observed radial dynamics can be compared with predictions based on Eq.~(\ref{eq:depth}) and the measured initial conditions. From the initial wave velocity ($v_{1}=\dot{\bar{r}}_{c} \approx 0.3\,\rm{\mu m/ms}$) and ring radius $\bar{r}_{{\mathrm{c}}}\approx 11~\mu$m measured at $t_{i}\approx 13\,\rm{ms}$, the dark ring is expected to reach a minimum radius $r_{\rm{min}} \approx 3\,\rm{\mu m}$, agreeing well with our observation approximately $3.2\,\rm{\mu m}$. To compare the entire density evolution with expectations, we fit the detected ring density dips [Fig.~\ref{fig:fig2}(c)] with $n(x) = |\psi(x,t)|^2 $ and extract the width $w$ as well as depth $d$ versus position of the defect center $x_{c}$ in panel (d). We compare the relationship $\xi \approx w\sqrt{d}$ in panel (e). The results are consistent with predictions assuming a perfectly unperturbed RDS, except that the measured depth stops increasing with decreasing ring size at a radius of less than or around $5~\mu$m. We attribute this reduced contrast to an instability developing in the dark ring, as we now discuss.

An RDS becomes unstable when the rotational symmetry is broken \cite{toikka2013snake}, which, in experiment, occurs in the presence of thermal and quantum fluctuations or with an azimuthal variation in the generating box potential. An RDS would suffer transverse modulations from self-amplifying noise. This is clearly visible in our experiments especially when the dark ring reaches the minimum radius, as seen in Fig.~\ref{fig:fig1}(d). By seeding initial fluctuations in an otherwise smooth GPE wave function (Appendix~\ref{App:GPE}), qualitatively similar density perturbations are observed in Fig.~\ref{fig:fig1}(c).

Interestingly, TI is intrinsically coupled to an RDS's radial motion. For a dark soliton stripe with finite length $L$, it is known that TI manifests as sinusoidal ``snaking'' density modulations along the stripe, with discrete wave numbers $k_l= 2\pi l/L\lesssim 1/w $ ($l \in \mathbb{N}$) limited by the transverse width $w$ \cite{ma2010controlling}. In an RDS, this sets a radius-dependent limit,
\begin{equation}
l \lesssim \frac{r_c}{w} \sim \frac{\sqrt[3]{r_\mathrm{min}r_\mathrm{c}^2}}{\xi}\quad \mathrm{for} \quad  l =1,2,3,... \,.\label{eq:bound}
\end{equation}
Therefore, high-frequency modulations stop growing as an RDS shrinks. Around the minimum radius, only the most unstable mode(s) with $l=l_\mathrm{max} \lesssim r_\mathrm{min}/\xi$ could continue to be amplified, until the RDS fragments into around $ l_\mathrm{max}$ pieces with angular separation $\Delta \phi \approx 2\pi /l_\mathrm{max}$.

\begin{figure}[!t]
\centering
\includegraphics[width=0.5\textwidth]{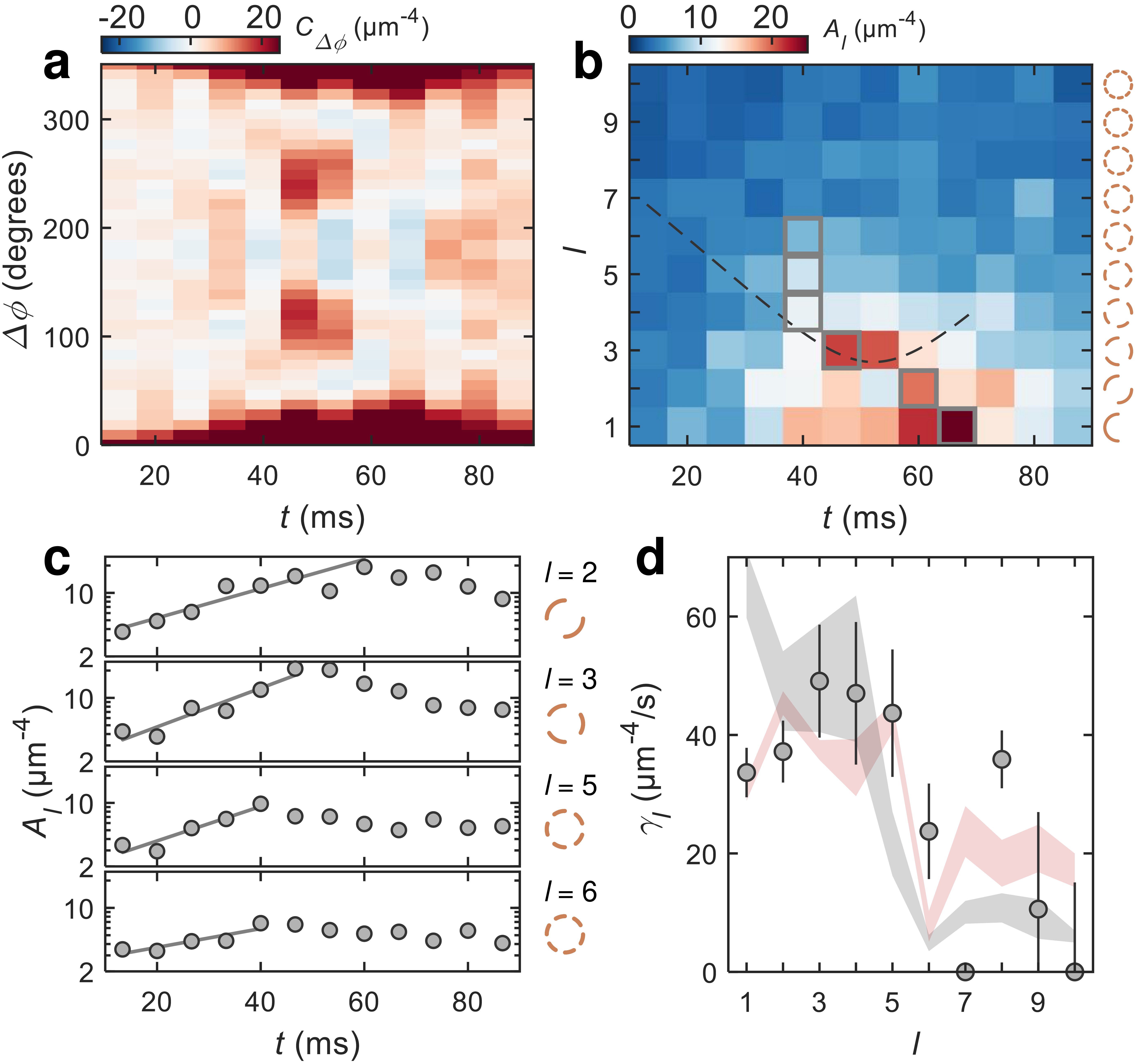}
\caption{
Pattern-forming instability. (a) Evolution of the azimuthal density-density correlation function $C_{\Delta \phi}$ showing pattern formation. (b) Fourier spectra $A_{l}$ showing mode competition. The peak position of each mode ($l\leq 6$) is marked by a gray square. The dashed line marks the calculated thresholds, Eq.~(\ref{eq:bound}), below which modes are expected to be unstable. (c) Fourier amplitude $A_{l}$ of indicated modes plotted in the logarithmic scale. Solid lines are exponential fits to determine the initial growth rates $\gamma_{l}$ shown in (d). 
The red shaded band represents the GPE simulated rate, scaled by an overall constant of around $0.17$ to match the data. The gray shaded band includes systematic effects in imaging, showing agreement with experiment without any adjustable parameters. Details can be found in Appendix~\ref{App:pattern formation GPE}. Error bars in the data and vertical bands in the simulated rate represent fitting uncertainty.
}
\label{fig:fig3}
\end{figure}

\begin{figure}[!t]
\centering
\includegraphics[width=0.5\textwidth]{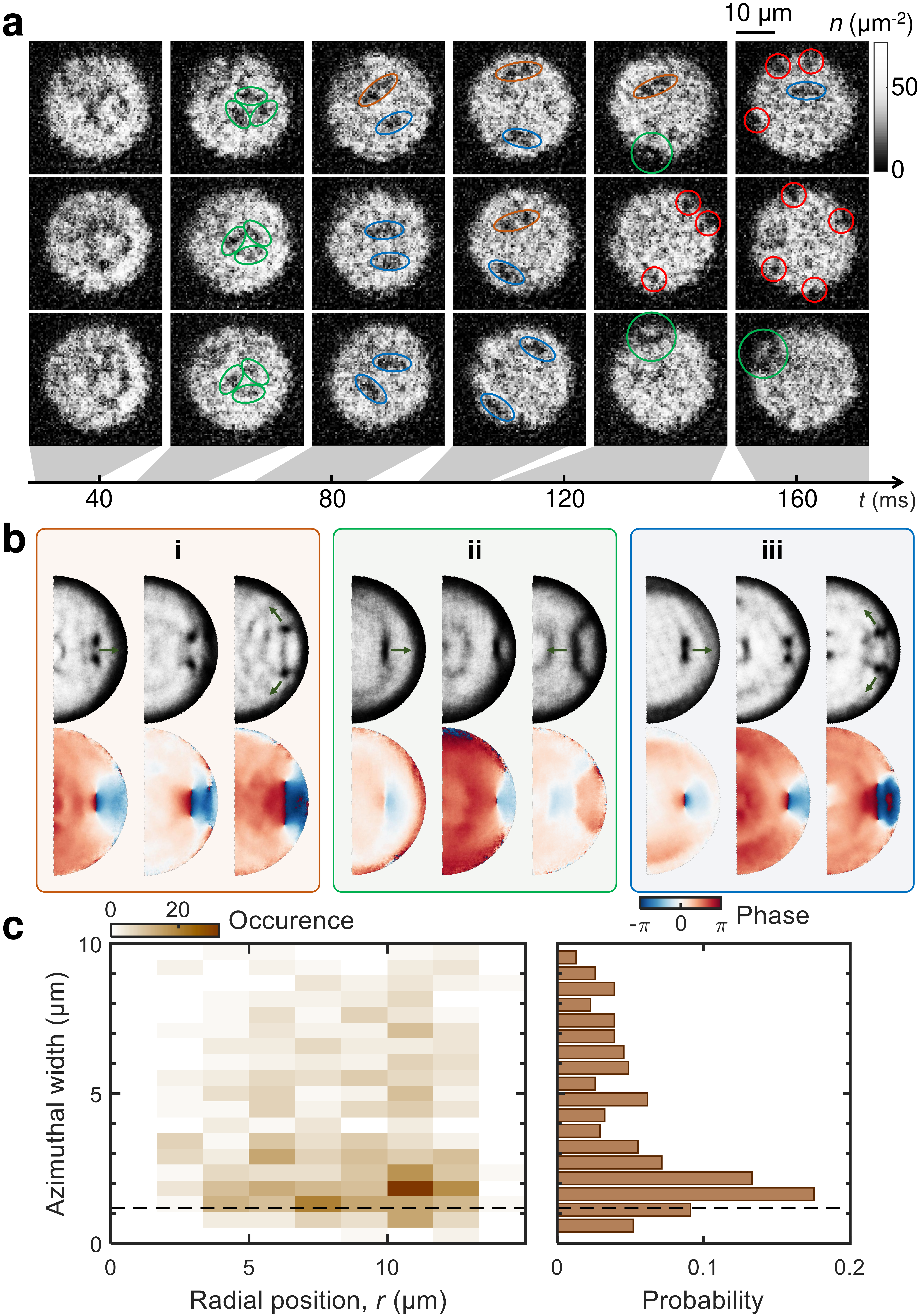}
\caption{Structured formation of vortex dipole necklace and vortex unbinding. (a) Single-shot in-situ images collected within the indicated time interval. The first three columns show samples right before, during, and right after patterned fragmentation, respectively. The last three columns present samples with density defects likely before, during, and after they reach the superfluid boundary.
Identified density defects are circled and categorized as one of the following: weakly bound vortex dipoles (brown), rarefaction pulses (green), bounded dipoles (blue), and pinned vortices (red). 
(b) Dynamics of a weakly bound dipole (i), a rarefaction pulse (ii), and a bounded dipole (iii) in GPE simulations. Images in each box, from left to right, respectively, show the density (top) and phase (bottom) profiles before and after a defect hits the wall. The propagation direction of each defect is marked by an arrow.
(c) Left: occurrence of azimuthal width versus radial position of detected defects, obtained from images taken after $t=60\,\rm{ms}$. Right: probability distribution for detected widths at $r\geq 9\,\rm{\mu m}$ showing a peak near the healing length (dashed line).
}
\label{fig:fig4}
\end{figure}

To visualize this dynamics, in Fig. \ref{fig:fig3}(a) we plot the angular density-density correlation function in the dark ring, $C_{\Delta \phi}=\Braket{C (\phi,\,\Delta \phi)}_{\phi}=\Braket{ \braket{n_{\phi} n_{\phi + \Delta \phi}} - \braket{n_{\phi}} \braket{n_{\phi + \Delta \phi}} }_{\phi}$, where $n_{\phi}=n(r_\mathrm{c},\phi )$ is the atomic density near the measured ring position $r = r_c$, and $\braket{\cdot}$ ($\braket{\cdot}_{\phi}$) denotes ensemble (azimuthal) averaging. After a long enough time while the dark ring approaches its minimum radius, we observe very strong angular correlations at $\Delta \phi \approx 120^{\circ}$ and $240^{\circ}$ angles, corresponding to $l=3$. This mode appears to create radial distortions of a triangular shape, as exemplified by the single-shot images in the second column of Fig.~\ref{fig:fig4}(a). Interestingly, as the dark ring continues to evolve ($t\gtrsim 50\,\rm{ms}$), a new correlation pattern develops at around $180^{\circ}$ angle, that is, for $l\leq 2$; see the images in the third column of Fig.~\ref{fig:fig4}(a).

To observe this mode competition more clearly, we measure the Fourier spectrum by evaluating $A_l = \Braket{\left|\sum_{\Delta \phi} C(\phi,\Delta \phi) e^{i l \Delta \phi}\right|}_{\phi}$. As shown in Fig.~\ref{fig:fig3}(b), instability develops mostly within $l\leq 6$, whose amplitudes are exponentially amplified as shown in panel (c). Figure 3(d) plots the initial growth rate for each mode, quantitatively reproduced by the GPE simulations detailed in Appendix~\ref{App:pattern formation GPE}. In the experiments, the $l=3$ mode is the most unstable with the largest growth rate.  At $t \gtrsim 40$~ms, however, the growth of high-frequency modes becomes arrested by the shrinking radius. The onset of growth suppression roughly follows the estimation given by Eq.~(\ref{eq:bound}), until the minimum ring radius is reached; see panel (b). Beyond $t\gtrsim 50~$ms, $l=1$ and 2 modes continue to increase until $t\gtrsim 70\,\rm{ms}$, when $A_{l=1,2}$ becomes large enough to break the expanding dark ring. In real space, this corresponds to fragmentation of $l_\mathrm{max}= 2$ pieces with $180^{\circ}$ angular separation as evidenced in our observations.

We identify these self-structured fragments as vortex dipoles \cite{theocharis2003ring} in which the vortex-antivortex distance $\Delta$ is linked to the speed $v_\mathrm{d} \sim \hbar/m \Delta$ and the incompressible kinetic energy of the flow $ E \sim \log (\Delta / \xi)$. Remarkably, we observe a variety of vortex dipole structures in experiment and in GPE simulations as well [Fig.~\ref{fig:fig4}(b)]; many of these structures were classified in Refs.~\cite{jones1982motions,jones1986motions}. One type of defect is referred to as a weakly bound vortex dipole, appearing when the flow has a larger energy. It features two well-separated cores and phase singularities as shown in (i).
A second type of defect is the rarefaction pulse shown in (ii), which shows a phase step without vorticities and propagates at a velocity closer to the sound speed. It emerges as a local density minimum weakly connected to other defects in the bulk or at the boundary. A third type, which we observe most frequently, is somewhat in between the first two. Its energy is large enough to preserve two phase vorticities but too small to separate their cores.
It can be seen as an isolated density defect with an elongated width ($\gtrsim 2w$) and is identified as a bounded dipole as in (iii). Oftentimes, the second and the third types of defects are referred to as coalesced vortices \cite{kwon2014relaxation} or Jones-Roberts solitons \cite{meyer2017observation}.

In Fig.~\ref{fig:fig4}(a), due to shot-to-shot fluctuations in atomic density (see also Appendix~\ref{App:vortex_algorithm}), observed density defects exhibit various shapes as discussed above.
Three rarefaction pulses (green ovals) that are weakly linked in a triangular shape are often found at $t\sim 50\,\rm{ms}$. Pairs of bounded vortex dipoles (blue ovals) are most frequently identified immediately after an RDS fragments.
Weakly bound dipoles (brown ovals) sometimes appear, potentially due to stronger snaking modulations and a larger flow energy to separate the cores.
At longer times, many isolated vortices (red circles) are found near the edge of the box, presumably exhibiting similar orbital dynamics as seen in Ref.~\cite{neely2010observation}. Detailed identification algorithms of density defects can be found in Appendixes~\ref{App:sol_algorithm} and \ref{App:vortex_algorithm}.

\begin{figure}[t!]
\centering
\includegraphics[width=0.5\textwidth]{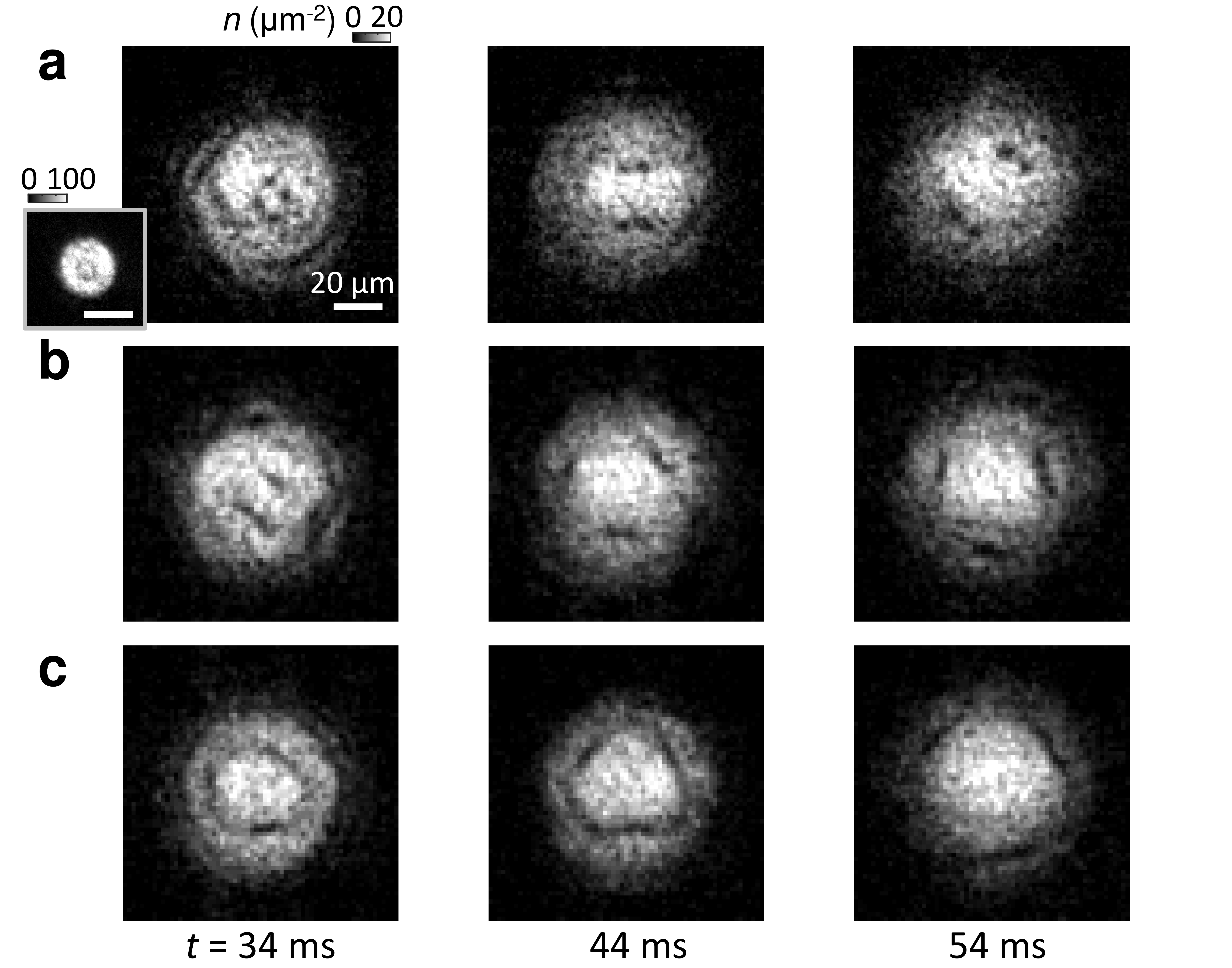}
\caption{Self-patterned defect structures imaged after $40\,\rm{ms}$ of TOF. Single-shot images in each column are collected with the indicated hold time, showing a necklace of weakly bound vortex dipoles (a), bounded dipoles (b), and rarefaction pulses (c). The inset shows a sample in situ image at $t=34\,\rm{ms}$, held in a circular box with a radius of approximately $ 11 \rm{\mu m}$. 
}
\label{fig:fig5}
\end{figure}

A clear distinction of vortex dipoles can be visualized from their interaction with the wall potential, where the density gradient induces an inward force on both the vortex and the antivortex and this triggers the Magnus effect. A vortex dipole would decelerate and unbind upon hitting the wall as shown in (i) and (iii) of Fig.~\ref{fig:fig4}(b). The unbound vortices appear pinned to the superfluid boundary and move along the rim with opposite circulation. If, instead, a rarefaction pulse is incident upon the wall, it forms an arclike defect structure, as shown in (ii), eventually breaking into a reflecting rarefaction pulse, and occasionally also two vortices pinned to the boundary with opposite circulations. These rarefaction arcs are evidenced by defects marked with large green circles in Fig.~\ref{fig:fig4}(a). Long after all vortex dipoles have interacted with the wall, we identify a high probability for observing vortexlike density defects (azimuthal size $\sim \xi$) near the boundary $r\gtrsim 9~\mu$m, as shown in Fig.~\ref{fig:fig4}(c). Other defects at $r\lesssim 9~\mu$m have a wide spread of azimuthal widths ($\gtrsim 2\xi$). They are likely rarefaction pulses, rebounded dipoles, or colliding defects in the bulk.

We have demonstrated \textit{in situ} images of self-structured density defects. This enables directly probing self-patterning dynamics in a superfluid for the first time. 
To further enhance visualization of vortex dipole necklaces, we can extinguish the horizontal trap confinement and image after a long time-of-flight (TOF) expansion in 2D; see Fig.~\ref{fig:fig5}. We note that these structures continue to evolve during the expansion.
Here, we use a smaller box to increase the initial density, chemical potential, and thus the TOF expansion rate.
Vortex cores in $l=2$ vortex dipole necklaces dramatically expand during TOF due to much-reduced healing length, clearly visible in panel (a).
As dipoles propagate towards the boundary, their core size further increases due to reduced background density.
We have also observed $l=2$ or $3$ bounded dipoles (b) and rarefaction pulses (c), identified based on their widths and connections with adjacent defects. Faster moving RDSs can also be seen near the boundary of expanded superfluids.

In summary, we observed very rich nonequilibrium dynamics and self-patterning with RDSs that emerged purely from a box quench. Both $180^\circ$ ($l_\mathrm{max} = 2$) and $120^\circ$ ordered ($l_\mathrm{max} = 3$) vortex dipole necklaces have been observed in Figs.~\ref{fig:fig4} and \ref{fig:fig5}, respectively. Even higher-order $l_\mathrm{max} \geq 3$ can be created with either a larger bulk chemical potential (for shorter-scale most unstable modes) or larger initial short-scale density perturbations (Appendix~\ref{App:GPE}). Our quench experiment demonstrates a new tool to generate dark solitons in versatile forms as well as ordered quantum vortex matter in a uniform box trap \cite{navon2021quantum}. By incorporating box quenches together with interaction tuning using a Feshbach resonance, multiple RDSs \cite{du2022quench,niu2023dynamical} and vortex dipole necklaces may be generated in one sample, thus creating complex vortex matter. It may be possible to trap a stationary RDS \cite{theocharis2003ring} and further control its stability by applying a radial potential when $r_\mathrm{c}=r_\mathrm{min}$ \cite{ma2010controlling,wang2015stabilization}. By incorporating nondestructive imaging \cite{serafini2017vortex}, our work can be extended to study inverted TI \cite{toikka2013snake}, persistent revivals and clusterization \cite{crasovan2002globally,mott2005stationary,pietil2006stability,cawte2021neutral} of ordered vortex dipoles, and may open new ways to explore spontaneous clustering \cite{simula2014emergence,billiam2014onsager,kanai2021true} in 2D vortex matter.

\section*{Acknowledgments}
We thank Eli Halperin, Qi Zhou, Samuel Alperin, Chih-Chun Chien, and Sergei Khlebnikov for discussions. This work is supported in part by the W. M. Keck Foundation, the NSF (Grant \# PHY-1848316), and the DOE QuantISED program through the Fermilab Quantum Consortium.

\appendix
\section{Preparation of a 2D superfluid} \label{App:prep}
The detailed experimental apparatus is given in Ref. \cite{chen2020observation} with an updated objective lens (numerical aperture of approximately $0.6$).
We begin the preparation of a Bose-Einstein condensate (BEC) of cesium atoms confined in an optical dipole trap with a horizontal (vertical) trap frequency of about $ 12\,\rm{Hz}$ (about $ 70\,\rm{Hz}$) through an evaporative cooling procedure.
The $s$-wave scattering length is then gradually decreased to a small value $a\approx12a_{0}$ via a Feshbach resonance \cite{chin2010feshbach}, where $a_{0}$ is the Bohr radius. The BEC is then loaded into a 2D box potential. The vertical confinement of the box is provided by a single node of a repulsive standing-wave potential with $ 3\,\rm{\mu m}$ periodicity. The measured vertical trap frequency in the node is $\omega_{z}\approx 2\pi \times 1.8\,\rm{kHz} $ ($\gg k_{B}T/\hbar,\, \mu /\hbar$), deep in the 2D regime, where $k_B$ is the Boltzmann constant, $T < 10~$nK is the temperature, $\mu$ is the chemical potential, and $\hbar$ the reduced Plank constant. The atoms populate the vibrational ground state in the vertical trap with a harmonic oscillator length $l_{z}=\sqrt{\hbar/(m\omega_{z})}\approx 207\,\rm{nm}$, where $m$ is the atomic mass. The horizontal confinement is arbitrarily configured via a blue-detuned light ($780\,\rm{nm}$) patterned with a digital mirror device and projected through the objective lens. In this work, we use a circular box with an inner radius of approximately $15 \,\rm{\mu m}$ and a width of approximately $5\,\rm{\mu m}$.
We obtain in situ density distributions of 2D gases by performing absorption imaging through the same objective lens and recording the image on a CCD camera. The image resolution is about $0.8\,\rm{\mu m}$. The atomic surface density $n$ is calibrated using a similar scheme as discussed in Ref. \cite{hung2011situ}. The typical initial density is $n \approx 50\,\rm{\mu m^{-2}}$ at a fixed 2D interaction strength $g=\sqrt{8\pi}a/l_{z}\approx0.017$. Shortly after the box potential height is quenched to the final strength ($\approx k_B \times 9~$nK), the bulk density reduces to $n \approx 42\,\rm{\mu m^{-2}}$ and remains roughly constant throughout the subsequent evolution, presumably due to the initial finite atom spilling over the box wall. The resulting healing length is $\xi  = 1/\sqrt{ng} \approx 1.2\,\rm{\mu m}$, larger than our image resolution.
This allows us to resolve individual vortices with spacing comparable to or smaller than the healing length. Note that in situ imaging of vortices has been demonstrated using a dark-field imaging technique \cite{wilson2015situ} and, most recently, using high-resolution absorption imaging \cite{kwon2021sound,fletcher2021geometric}.

 \begin{figure*}[t]
\centering
\includegraphics[width=0.8\textwidth]{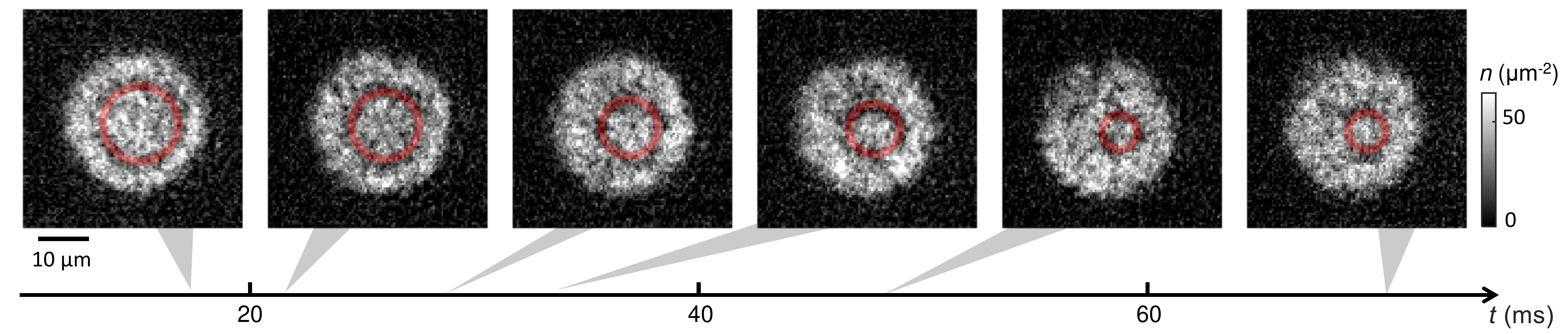}
\caption{Single-shot in-situ images of RDSs. The red shaded area represents the radius $\bar{r}_{c}$ of produced RDSs from the averaged image. Even though RDSs fragment into multiple pieces for a longer time, their locations are efficiently captured.
}
\label{fig:figSM1}
\end{figure*}

\section{Wave function and radial dynamics of a ring dark soliton}\label{App:radial}
An RDS is a quasistationary solution of the time-dependent 2D Gross-Pitaevskii equation (GPE),
\begin{equation}
    i \hbar \frac{\partial  \psi}{\partial t} = \left[-\frac{\hbar^2}{2m}\left( \frac{\partial^2}{\partial r^2} + \frac{1}{r} \frac{\partial }{\partial r} + \frac{1}{r^2}\frac{\partial^2}{\partial \phi^2}\right) + \frac{\hbar^2 g}{m} |\psi|^2\right] \psi \, .
\end{equation}
Assuming rotational symmetry, the wave function of a perfect RDS only has radial dependence. To good approximation, it can be written as 
\begin{equation}
    \psi(r,t) =\sqrt{n}\left[ i\sqrt{1-d} + \sqrt{d} \tanh (r-r_\mathrm{c})/w \right] e^{-i \mu t/\hbar} \, ,
\end{equation} 
where $r_\mathrm{c}(t)$ is the radial position of the density defect, $d=1-(v/v_\mathrm{s})^2 \leq 1$ its depth, $v=\dot{r}_\mathrm{c}$ its radial velocity, and $w=\xi/\sqrt{d}$ its characteristic width. Here, the background density $n$, the healing length $\xi = 1/\sqrt{ng}$, the sound speed $v_\mathrm{s}=\hbar/\xi m$, and the chemical potential $\mu =  m v_\mathrm{s}^2$ are the bulk properties of the superfluid. The radial motion $v$, width $w$, and depth $d$ are all related to each other; fixing one parameter completely determines the other two. A faster-(slower-)moving soliton would have a shallower (deeper) depth and a broader (narrower) density profile.

This radial wave function is essentially the dark soliton solution in 1D, except that it is perturbed by the $r^{-1}\partial/\partial r$ Laplace term in the 2D GPE. A consequence of this perturbation is that $(d,w,v)$ slowly evolves as the radius of an RDS changes \cite{kivshar1994ring}.
The dynamics of an RDS differs from that of a 1D dark soliton. In particular, the soliton depth follows the relation 
\begin{equation}
    d = d(t_{i}) \left[ \frac{r_\mathrm{c}(t_{i})}{r_\mathrm{c}} \right]^{2/3} \, ,    
\end{equation}
where $d(t_{i})$ and $r_c(t_{i})$ are the initial depth and radius of the RDS, determined at time $t_{i}$. This additional equation further relates $(d, w, v)$ with $r_\mathrm{c}$. The depth increases (decreases) as the RDS shrinks (expands), and the width and radial velocity change accordingly. 

An explanation for this radius-dependent dynamics is from energy conservation. As discussed in Ref. \cite{brazhnyi2006dark}, the energy of a 1D dark soliton is $\epsilon=(4/3)\hbar v_\mathrm{s} n d^{3/2}$. For a dark soliton stripe in 2D, $\epsilon$ is the linear energy density. In a uniform medium, the total energy of an RDS is $2\pi r_{c} \epsilon$. For an adiabatic evolution with conserved RDS total energy, one must have $r_{c}(t_{i}) d(t_{i})^{3/2} = r_{c} d^{3/2}$, thus leading to the same result obtained from the perturbation theory \cite{kivshar1994ring}.

\begin{figure*}[!t]
\centering
\includegraphics[width=0.8\textwidth]{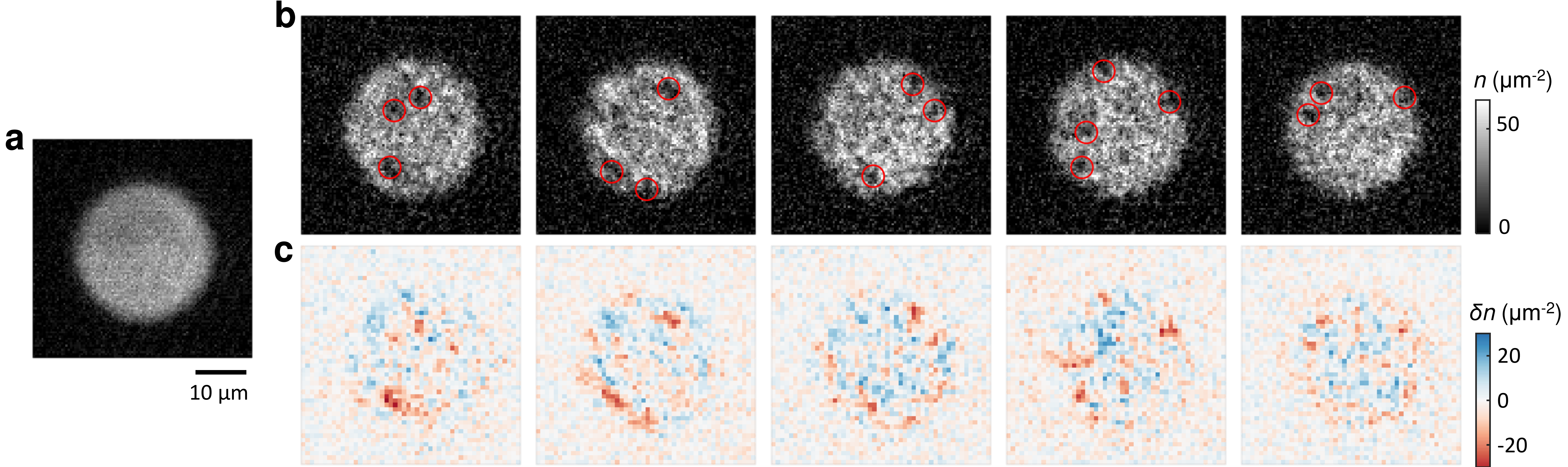}
\caption{In situ vortex detection in a box-trapped 2D superfluid. (a) Mean density profile $\bar{n}$ averaged over more than $300$ experimental realizations at over $60\,\rm{ms}$. (b) Examples of single-shot in-situ density images $n$. Detected defects are marked by red circles. (c) Residual from the mean density, $\delta n = n - \bar{n}$. 
}
\label{fig:figSM2}
\end{figure*}

\section{Time-dependent GPE simulation}\label{App:GPE}
We perform 2D GPE simulations \cite{antoine2014gpelab,antoine2015gpelab} to obtain numerical evidence of RDS emission in our quench protocol. The initial ground-state wave function is confined in a repulsive wall potential of the form
\begin{equation}
    U(r) = \begin{cases}
      U_0 e^{-2 (r-R)^2/\sigma^2} & r \leq R \\
      U_0 & r >R\\
    \end{cases}   
\end{equation}
where $U_0$ is the trap strength, $R$ the box radius, and $\sigma$ the experimentally calibrated $1/e^2$ width of the wall. In the time-dependent GPE, the trap strength is quenched from $U_0=k_B\times 60~$nK to the final value $U_f=k_B\times 9~$nK, and we evaluate the subsequent dynamics of the wave function. We note that, in the experiment, the wall has a radial Gaussian profile of finite width instead of the semi-infinite form taken in the GPE simulation. The former is responsible for finite atom spilling after the potential is quenched down. The results shown in Fig.~\ref{fig:fig1} are calculated using atom number $N=2.2 \times 10^4$ matching that of the initial experimental condition. While we have obtained qualitative one-to-one agreement of RDS emission and its subsequent evolution, the exact timing cannot be fully matched. The emitted RDS tends to move faster in the GPE simulation, and we have increased $U_0$ by about $ 70\%$ to increase the quench contrast, which slows down the RDS velocity. The slower RDS dynamics observed in the experiment may be due to finite atom spilling right after the quench, which leads to a lower sound speed and thus a lower RDS velocity.

To take into account density fluctuations in a superfluid, we imprint phase noise in the initial GPE wave function to simulate phonon excitations. Given an initial temperature ($T\approx 3-7~$nK), we calculate the phonon populations according to the Bose-Einstein distribution plus zero-point fluctuations,
\begin{equation}
  n_p(k)=\frac{1}{e^{E(k)/k_B T}-1} + \frac{1}{2} = \frac{1}{2}\coth \frac{E(k)}{2k_B T}  \, ,
\end{equation}
where $E(k)$ is the Bogoliubov phonon dispersion relation. We populate random Bogoliubov phonon excitations in the ground-state wavefunction, with statistical amplitude variance in each mode matching $n_p(k)$. We then evolve the wave function in the time-dependent GPE. We have also taken into account total atom number fluctuations in the experiment, and performed a series of GPE calculations with $N=(2.2 \pm 0.4 )\times 10^4$. Given the range of temperature and atom number fluctuations, we numerically observe RDS fragmentation into necklaces of weakly bound vortex dipoles, tightly bound dipoles, or rarefaction pulses. Representative results are plotted in Fig.~\ref{fig:fig4}(b). Most of the necklaces consist of a chain of $l=2$ vortex dipoles or rarefaction pulses. Increasing the atom number or interaction strength $g$, $l \geq 3$ necklaces can be observed.

\section{Soliton detection algorithm}\label{App:sol_algorithm}
To obtain the position and size of an RDS, we measure mean density line cuts $\bar{n} (x,\,y_{0})$ typically averaged over realizations equal to or greater than 5. By convolving the line cuts with a Gaussian kernel of a width around the healing length $\xi$, we obtain a scale-space representation that suppresses features smaller than $\xi$. A Laplacian of this convolution generates positive features for intensity minima, which correspond to the mean locations of dark stripes. These procedures are applied to the density line cuts obtained at different times $t$ to visualize the evolution of dark stripe locations, as shown in the right panel of Fig.\ref{fig:fig2}(a). We then extract the stripe speed and location. 
We perform the same analysis to averaged density line cuts along the $y$ axis, $\bar{n} (x_{0},\,y)$.
With this information obtained from the mean line cuts along the two axes, we predict the time-dependent center $r_{0}$ and radius $r_{c}$ of RDSs in single-shot images; see red areas in Fig.~\ref{fig:figSM1}.
For the analysis of instability in RDSs (see Fig. 3) we use single-shot density profiles averaged in the radial interval $|r - r_{c}| \lesssim  0.9\,\rm{\mu m}$.

\section{Vortex detection algorithm}\label{App:vortex_algorithm}
At later times ($t\gtrsim 60\,\rm{ms}$) after the box quench, an RDS breaks into vortices. They are detected using a scheme similar to a vortex image processing algorithm \cite{rakonjac2016measuring}.
For each single-shot in situ image [Fig.~\ref{fig:figSM2}(b)], we calculate the residual [Fig.~\ref{fig:figSM2}(c)] from the mean density profile [Fig.~\ref{fig:figSM2}(a)] averaged over many experimental shots. We then apply a Laplacian of the Gaussian filter to enhance those defect structures having characteristic length scales of about $\xi$. Then, we obtain defect positions marked by circles in Fig.~\ref{fig:figSM1}(b). We note that the analysis is restricted to defects having nearly zero local density ($\lesssim 4\,\rm{\mu m^{-2}}$) to avoid spurious detections. Detected defects have a large variation in the azimuthal width, while the radial spread is comparable with around $\xi$.
The azimuthal width and position of each detected defect are analyzed as shown in Fig. 4(c).

\begin{figure*}[t]
\centering
\includegraphics[width=\textwidth]{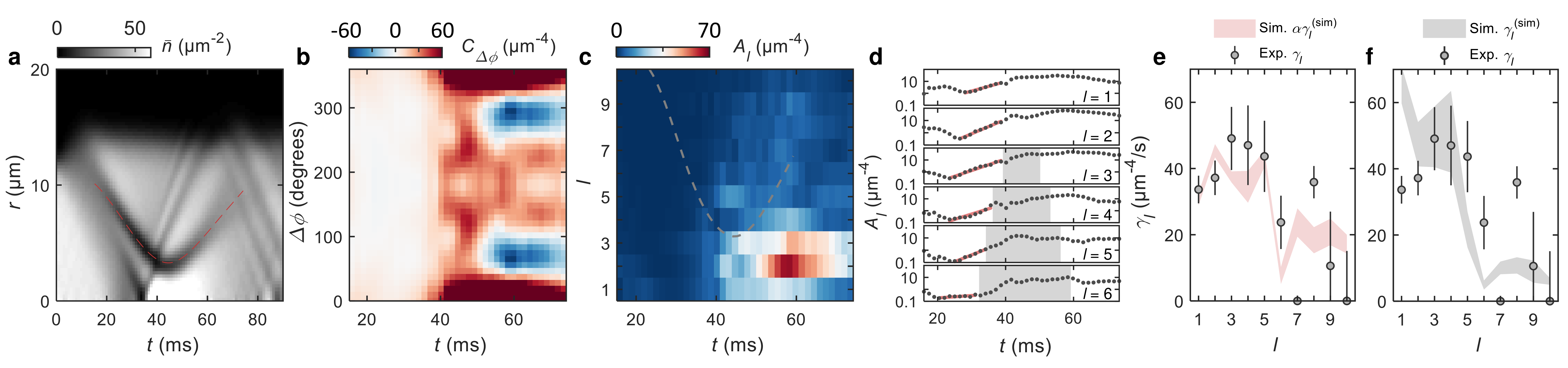}
\caption{Pattern-forming instability in the simulation. (a) Radial density averaged over nine runs. Dashed line represents detected radius $\bar{r}_{c}$ of generated RDSs. (b) Azimuthal density-density correlation function $C_{\Delta \phi}$. (c) Fourier spectra $A_{l}$. Dashed line shows the maximal unstable mode $l_{\rm{max}}$, expected from $\bar{r}_{c}$ and $\dot{\bar{r}}_{c}$. (d) Dynamics of $A_{l}$ in the logarithmic scale. Solid lines are exponential fits to determine the initial growth rates $\gamma_{l}^{(\rm{sim})}$. Shaded area shows a regime where the mode amplification is expected to be halted $l>l_{\rm{max}}$.   (e) Experimental rates $\gamma_{l}$ (dots), compared with $\gamma_{l}^{(\rm{sim})}$ scaled by an overall factor $\alpha\approx 0.17$ (shaded area). Error bars and vertical width in $\alpha \gamma_{l}^{(\rm{sim})}$ represent fitting uncertainty. (f) Same as (e), except that $\gamma_{l}^{(\rm{sim})}$ is from the simulations including systematic effects. The simulated rates well reproduce the data with no free parameters.
}
\label{fig:figSM3}
\end{figure*}

\section{Pattern formation dynamics in GPE simulation}\label{App:pattern formation GPE}

A series of GPE simulations with initially seeded random fluctuations are analyzed according to the presented data in Fig.~3 for further understanding the pattern formation by snaking instability in RDSs.
We first determine the mean radius $\bar{r}_{c}$ of generated RDSs and its motion $\dot{\bar{r}}_{c}$ from a radial density profile [Fig.~\ref{fig:figSM3}(a)], averaged over multiple runs. We then analyze the angular density-density correlation function $C_{\Delta \phi}=\Braket{C (\phi,\,\Delta \phi)}_{\phi}=\Braket{ \braket{n_{\phi} n_{\phi + \Delta \phi}} - \braket{n_{\phi}} \braket{n_{\phi + \Delta \phi}} }_{\phi}$, where $\braket{\cdot}$ ($\braket{\cdot}_{\phi}$) represents the sample (azimuthal) average and $n_{\phi}$ is the azimuthal density at $r=\bar{r}_{c}$. 
As shown in Fig.~\ref{fig:figSM3}(b), the simulated $C_{\Delta \phi}$ is consistent with the observation [Fig. 3(a) in the main text]. 
The RDSs have no significant modulation in their azimuthal density at early times, less than or around $20\,\rm{ms}$. However, at a later time when $\bar{r}_{c}$ approaches its minimal value, $C_{\Delta \phi}$ exhibits strong angular correlation at $\Delta \phi \approx 120^{\circ}$ and $240^{\circ}$, corresponding $l=3$ necklace formation. At even later times, the angular correlation at $\Delta \phi \approx 180^{\circ}$ ($l=2$ mode) becomes comparable to the ones at $\Delta \phi \approx 120^{\circ}$ and $240^{\circ}$, resulting in the competition among those unstable modes.

In Fig.~\ref{fig:figSM3}(c), we plot the dynamics of the Fourier spectrum, defined as $A_l = \Braket{\left|\sum_{\Delta \phi} C(\phi,\Delta \phi) e^{i l \Delta \phi}\right|}_{\phi}$. It marks the increase of the modulation amplitude only for lower angular frequency modes due to the finite instability band $l<l_{\rm{max}}$; the $l=3$ mode dominates at $t\sim40\,\rm{ms}$, while $l_{\rm{max}}$ approaches $r_{\rm{min}}/\xi \sim3$ as the RDS shrinks.
At $t\sim 60\,\rm{ms}$, the $l=2$ and $l=3$ modes eventually display comparable amplitudes. This dynamics naturally appears in shrinking RDSs, supporting our experimental observation of mode competition dynamics in Figs.~3 (b) and 3(c). 
The detailed dynamics of $A_{l}$ is plotted in Fig.~\ref{fig:figSM3}(d). 
We find that the amplifications in the prohibited band $l>l_{\rm{max}}$ are incompletely terminated at $l\geq 4$. This can potentially be attributed to the contribution from adjacent RDSs [see panel (a)] or the evolution of phonon fluctuations that may not be sensitively detected in the experiment.
Additionally, in the simulation, one can see a continuous reduction in $A_{l}$ at the early stage $t\lesssim 20 \,\rm{ms}$ during which the wave function expands while solitary waves develop. Note that $A_{l}$ starts to grow once RDSs form.
 
Figures \ref{fig:figSM3}(e) and \ref{fig:figSM3}(f) compare the experimental data $\gamma_{l}$, as shown in Fig.~3(d), with the early-time growth rate $\gamma_{l}^{(\rm{sim})}$, obtained from exponential fits as in Fig.~\ref{fig:figSM3}(d). In Fig.~\ref{fig:figSM3}(e), the observed exponential growth within $l\lesssim 6$ is well reproduced by the simulation, except for an overall scaling constant $\alpha\approx 0.17$ adjusted to match the data, indicating that the observed rate is roughly 6 times lower than that in the simulation. 
We believe the slower rate observed in the experiment is due to the systematic effect of finite resolution of our imaging system and image noise, for example, photon shot noise, which adds to each measured Fourier amplitude $A_l$ an offset of about $ O(1)$. 
We find that the offset can effectively reduce the fitted growth rates. To see those systematic effects, we convolute the GPE data with finite resolution and add the measured offsets to the GPE simulated amplitudes $A_l$. We refit the growth rate $\gamma_{l}^{(\rm{sim})}$ and indeed find that it agrees well with experimentally determined values, as shown in Fig.~\ref{fig:figSM3}(f).

\end{document}